# The role of focused laser plasmonics in shaping SERS spectra of molecules on nanostructured surfaces

Fran Nekvapil [a] and Cosmin Farcău [*ab]

Over fifty years have passed since the groundbreaking discovery of Surface Enhanced Raman Scattering (SERS), yet many aspects of this phenomenon remain elusive. In this study, we unveil novel observations concerning the spatial variation of SERS signal profiles through vertical ($Z$ axis) scans, performed by varying the distance between the laser focus and a solid, planar, nanostructured SERS substrate. The signal strength profile manifests a Lorentzian shape during axial scans along the $Z$ direction, consistently peaking above the actual sample surface. More intriguingly, the intensity ratio of various spectral regions—including SERS bands and background—exhibits significant non-constancy along the $Z$ axis. Finite-Difference Time-Domain (FDTD) simulations suggest that these variations can be attributed to specific plasmonic near-field responses induced by the focused/defocused beam at the SERS substrate. This research highlights the critical need to consider that focus imprecision can alter spectral profiles in SERS analyses on solid nanostructured SERS substrates, particularly when devising quantitative assays based on band intensity ratios.

## 1. Introduction

Surface-enhanced Raman spectroscopy (SERS) is a molecular vibration analytical tool, based on amplifying the Raman signal of interatomic bonds that interact with a plasmonic substrate. This amplification is primarily due to the enhancement of electromagnetic fields, owing to localized surface plasmon resonances. Several reviews are available detailing the SERS effect and the various SERS substrates realized with noble metals such as gold[1,2] and silver,[3] and in combination with semiconductors such as zinc oxide.[4] There are also reports of SERS on nanostructured platinum and palladium substrates, however, such elements work better in combination with gold or silver in core–shell or laminar structures.[5] Confocal Raman microscopy leverages the capability of spatially filtering information in the sample, enabling the creation of high-resolution chemical maps ($X$ and $Y$ directions) and depth profiles ($Z$ direction).[6] Confocality can be achieved using pinholes in the scattered beam path. Such measurement precision also comes with more danger of incorrect interpretation of results due to phenomena that become apparent, and relevant, only at this analytical scale.

One category of such issues can be related to proper or improper laser focusing. In a Raman micro-spectroscopy experiment, potential sources of imprecise control over the focus might be difficulties in visual or automatic adjustments on smooth, inhomogeneous or laminar samples. Furthermore, focusing based on video imaging does not always coincide with the position of the maximum collected signal. These may in turn influence the SERS signal that is recorded. The rationale for SERS determination of analyte concentration in mixtures is the proportionality between the amount of a particular atomic bond or molecular group within the probing area and the signal intensity. However, configuration of the bands belonging to the same compound could furthermore indicate its orientation relative to the SERS substrate. Indeed, several previous studies attempted to determine the analyte orientation *via* the ratio between different SERS bands.[7–9] If this principle is robust and precise, then the band configuration (band ratios) should be the same at any proximal point of the $Z$-axis, regardless of whether the analyte is in tight focus. As we show in the following, this may not be the case in certain situations. In this study, we explored the effect of the $Z$-axial position of the excitation beam focus on the SERS signal from nanostructured plasmonic surfaces using confocal Raman microscopy. This was achieved by recording and processing $Z$-linescans on an analyte–SERS substrate system consisting of 4-aminobenzenethiol molecules adsorbed on a gold-coated polystyrene nanosphere array. Two different objectives were used: one with a high numerical aperture (NA) and the other with a low NA, while the sample was moved below and above the objective focal point. The non-constant variation of various spectral regions of the SERS spectra were analysed. The experimental results were accompanied by finite-difference time-domain (FDTD) simulations of the electromagnetic field in the

[a] *National Institute for Research and Development of Isotopic and Molecular Technologies, Cluj-Napoca, Romania. E-mail: cfarcau@itim-cj.ro*
[b] *Institute for Interdisciplinary Research in Bio-Nano-Sciences, Babeș-Bolyai University, Cluj-Napoca, Romania*

proximity of the metal surface, and a plasmonic mechanism for the observed effects was proposed.

## 2. Materials and methods

### 2.1. SERS substrate fabrication and characterization

The first step in the fabrication of SERS substrates is the preparation of ordered arrays of colloidal polystyrene nanospheres on solid substrates. These colloidal arrays were prepared on rectangular polystyrene plates using a custom-made convective self-assembly equipment. The polystyrene plates (1 mm thick) were first cleaned with ethanol and isopropanol before drying by nitrogen blowing. The surfaces of these plates were rendered hydrophilic by a 20 min UV–ozone treatment, before being transferred and fixed on a motorized translation stage controlled through a computer interface. Water-based colloidal suspensions of 460 nm sized polystyrene spheres (10% w/v, Merck/Sigma-Aldrich) were used. A drop of colloidal suspension (∼10 μL) was inserted into the space between the substrate and the edge of a rectangular glass plate fixed near the substrate and inclined at an angle of ∼25°. Subsequently, by translating the substrate relative to the fixed plate, colloidal crystal films were formed on the substrate. Colloidal monolayers were obtained by setting the translation speed in the 45–60 μm s$^{-1}$ range. The experiments were performed under ambient temperature and humidity conditions. On top of the PS nanosphere arrays, gold films with a thickness of ∼150 nm (±8 nm) were deposited via e-beam evaporation (Kenosistec KE400) to obtain gold-coated microsphere arrays, sometimes also known as gold film over nanospheres (AuFoN) in the literature.[10–14] Scanning Electron Microscopy (SEM) images were obtained using a Hitachi SU8230 system. Optical reflectance spectra were recorded using a Technospex uSight-2100 microscope-mounted spectrophotometer.

### 2.2. SERS analysis

For SERS studies, 10$^{-5}$ M 4-aminobenzenethiol (4-ABT) (Sigma-Aldrich) in methanol solutions were prepared. The fabricated SERS substrates were immersed in the 4-ABT solution for 5 h, followed by rinsing with methanol. The Witec Raman spectrometer model Alpha300 was used to acquire linescans along the Z-axis. A typical linescan is composed of 25 spectra. The 632.8 nm laser (0.4 mW power) was used. The pinhole diameter was set to 25 μm. Two objectives with different numerical aperture (NA) were used: an 100×, high-NA (NA 0.9) and a 20×, low-NA (NA 0.4) objective. In general terms, the sampling volume is a spheroid with a beam waist defined by $D_{x,y} = 1.22\lambda/\text{NA}$, and a vertical circumference defined by $D_z = (4n\lambda)/(\text{NA})$, where $\lambda$ is the excitation wavelength, NA is the numerical aperture of the microscope objective and $n$ is the refractive index of the sample.[15,16] A circular laser spot with diameter ($D_{x,y}$) can thus be assumed on the SERS substrate at optimal focus of 0.86 and 1.93 μm, respectively. Linescans were obtained using an XYZ translation stage, and the movement ranges were set to approximately the focal depth of respective objectives determined using the formula mentioned above ($D_z(\text{NA } 0.9) = 3.12$ μm; $D_z(\text{NA } 0.4) = 15.82$ μm). Hence, in the case of the 0.9 NA objective, the z-range was 6 μm with a point spacing of 0.24 μm (from 3 μm above the sample plane to 3 μm below), whereas in the case of the 0.4 NA objective, the range was 30 μm with a point spacing of 1.2 μm (15 μm above and 15 μm below the sample plane).

To analyse the Raman intensity profiles along the linescans, four distinct spectral regions (50 cm$^{-1}$ wide) were chosen, and their integrated areas were plotted against the Z position. To plot the SERS strength profiles of 4-ABT signal, the background was subtracted from all spectra using an automatic function of the Witec Project FOUR software under the curvature parameter of 100. Subsequently the regions of the two main SERS bands were considered: the profile F1 comprises the band at 1080 cm$^{-1}$ and the profile F2 refers to the complex feature around 1580 cm$^{-1}$. Both of these bands are assigned to the phenyl ring breathing of 4-ABT which are presumably strongly coupled to the substrate, as will be discussed later. We also took advantage of the SERS background of unprocessed spectra to track the signal strength profile in two regions without the analyte-specific SERS bands. Hence, we selected the profile F3 referring to the region close to the laser wavelength, around the 280 cm$^{-1}$ wavenumbers, and profile F4 referring to the 2600 cm$^{-1}$ region far from the laser line. Converted to wavelengths, profiles F1, F2, F3 and F4 amount to 679, 703, 644 and 757 nm, respectively.

### 2.3. FDTD simulations

The optical/electromagnetic response of the AuFoN structure was computed using the FDTD method, with the aid of ANSYS Lumerical FDTD software. We designed a realistic structure to closely mimic the morphology of a real sample. A finite array of 90 polystyrene spheres was arranged in a close-packed hexagonal structure on a dielectric substrate. The spheres lattice was covered with a 150 nm thick gold film. Triangular metal nanoparticles formed on the substrate by projection through the openings between each three adjacent spheres were also modelled. The material properties included in the Lumerical database were used, which are based on Johnson and Christy[17] for Au. Refractive indices of 1.59 and 1.00 were used for polystyrene and air, respectively. All boundaries of the simulation region were surrounded by perfectly matched layer (PML) conditions. An additional mesh with a spatial resolution of 4.6 nm, was defined throughout the AuFoN, extending 50 nm above and below the AuFoN. A Gaussian source linearly polarized along the X direction was used to inject light along the Z direction from the sample side. Simulations were performed for different Z positions of the focus. Electric fields were mapped using field-profile monitors positioned perpendicular to the AuFoN in the XZ plane. More information on the fine details of the constructed structure can be found in a previous publication.[18]

## 3. Results and discussion

### 3.1. Characterization of the AuFoN SERS substrate

The AuFoN SERS-active substrate consists of polystyrene spheres (460 nm in diameter) coated with a gold film (150 nm

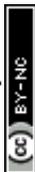

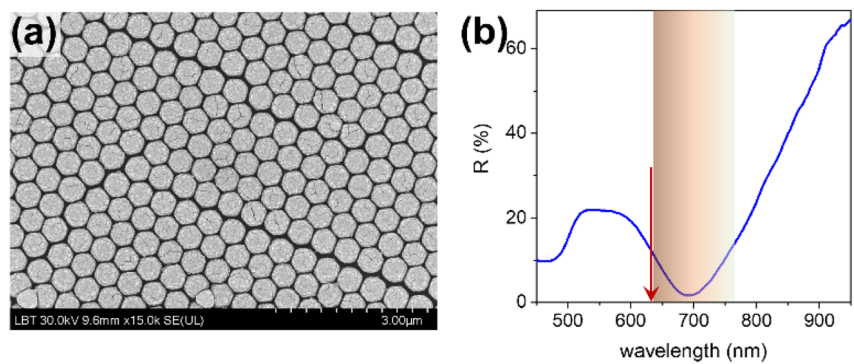

Fig. 1 (a) SEM image of AuFoN; (b) reflectance spectrum; arrow indicates 633 nm, used as excitation in SERS analyses, while the highlighted area depicts the spectral region of Raman scattered photons.

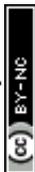

thickness). The SEM images confirm the dimensions of the nanoscale features, sphere packing, and some typical lattice defects usually found in this type of structure (Fig. 1a). Although such defects are present, the analyses in this study were done under a microscope, which allowed to visualize the sample surface and select uniform regions, which are larger than the lateral focus dimensions.

The most relevant feature in the measured reflectance is the broad dip centred around 690 nm, with increased reflectance values for shorter wavelengths and a steep increase towards longer wavelengths in the near-infrared region (Fig. 1b). The sphere size and gold film thickness were chosen based on previous experimental evidence that the position of this reflectance minimum indicates the greatest plasmonic activity of this type of SERS substrate, and that the strongest SERS enhancements are observed around this reflectivity minimum.[10–12] Our SERS substrate is thus very well adapted for excitation at 633 nm (arrow in Fig. 1b), offering high SERS enhancements in the range of 633–760 nm (corresponding to the Raman range up to 2600 $cm^{-1}$), marked by the shaded area in Fig. 1b.

### 3.2. Impact of objective numerical aperture on focusing/defocusing

Typical SERS spectra acquired in a Z-linescan are shown in Fig. 2, along with a scheme of the experimental setup. Fig. 2a schematically shows the layout of the axial Z-linescan points, while Fig. 2b shows the geometrical scheme of the laser beam propagation for the high-NA and low-NA objectives. In the current study, positive Z positions refer to the situation where the objective focus is placed above the sample plane, whereas negative Z positions refer to the focus below the sample plane.

4-ABT molecules are capable of forming self-assembled monolayers on gold surfaces through thiolate-metal bonds with the substrate. The van der Waals forces may further mediate the interactions between the tail groups of the molecules.[9,19,20] Previously reported SERS spectra of 4-ABT show that configuration and intensity of the bands seems to ultimately depend on the SERS substrate. The two dominant features in all recorded spectra are the band at 1080 $cm^{-1}$ and the complex feature around 1580 $cm^{-1}$, assigned to the C–S stretching and C–C stretching at the phenyl ring of 4-ABT,[19] indicating that the phenyl ring is strongly coupled with the AuFoN surface via the adsorption of the C–S group. According to SERS selection rules based on the electromagnetic propensity,[9,21] the information extracted from the spectra suggest that the phenyl ring does not presumably lie flat on the plasmonic surface but is rather perpendicular to it.

Fig. 2c shows an interesting observation when a high-NA objective was used. The spectrum recorded with the focus at the sample plane ($Z = 0.12$ μm) exhibits by far a stronger signal than the spectra equally spaced above and below the focus (z-positions −1.5 and 1.5 μm, respectively), which exhibit in turn a stronger signal than spectra at the ends of the linescan (z-positions −3 and 3 μm, respectively). Fig. 2e shows that the difference in both band intensity and background level occurs mainly in the 400–1600 $cm^{-1}$ range, which roughly corresponds to the wing of the reflectance behaviour from Fig. 1, from the laser wavelength towards the reflectance minimum around 690 nm. The difference was less pronounced in the higher-wavenumber region. In contrast, the Z-linescans recorded using the low-NA objective did not show such a pattern (Fig. 2d), with the constituent spectra being similar in terms of intensity (Fig. 2f), and thus further presentation of experimental results will refer only to the high-NA measurements.

### 3.3. SERS signal strength at different focus Z levels

The SERS signal strength variation along the Z-linescan (Fig. 3a) was determined by tracking the integrated areas of the spectral features (profiles F1, F2, F3, and F4). The Z-direction signal strength evolution profiles can be described quite well in all cases with Lorentzian peak shapes. The non-linear signal strength evolution is also permitted by Gaussian beam optics, where the laser spot diameter is the smallest at the beam waist, but increases non-linearly at any other beam transect. This is significantly influenced by the objective NA and the angles at which the incident and the scattered beams are handled by the objective. Hence, under a 0.9 NA objective the non-linearity of the spot size evolution along the z-direction, and thus the SERS signal variation is expected to be pronounced.[22,23] At the same time, previous reports on this kind of SERS substrates have

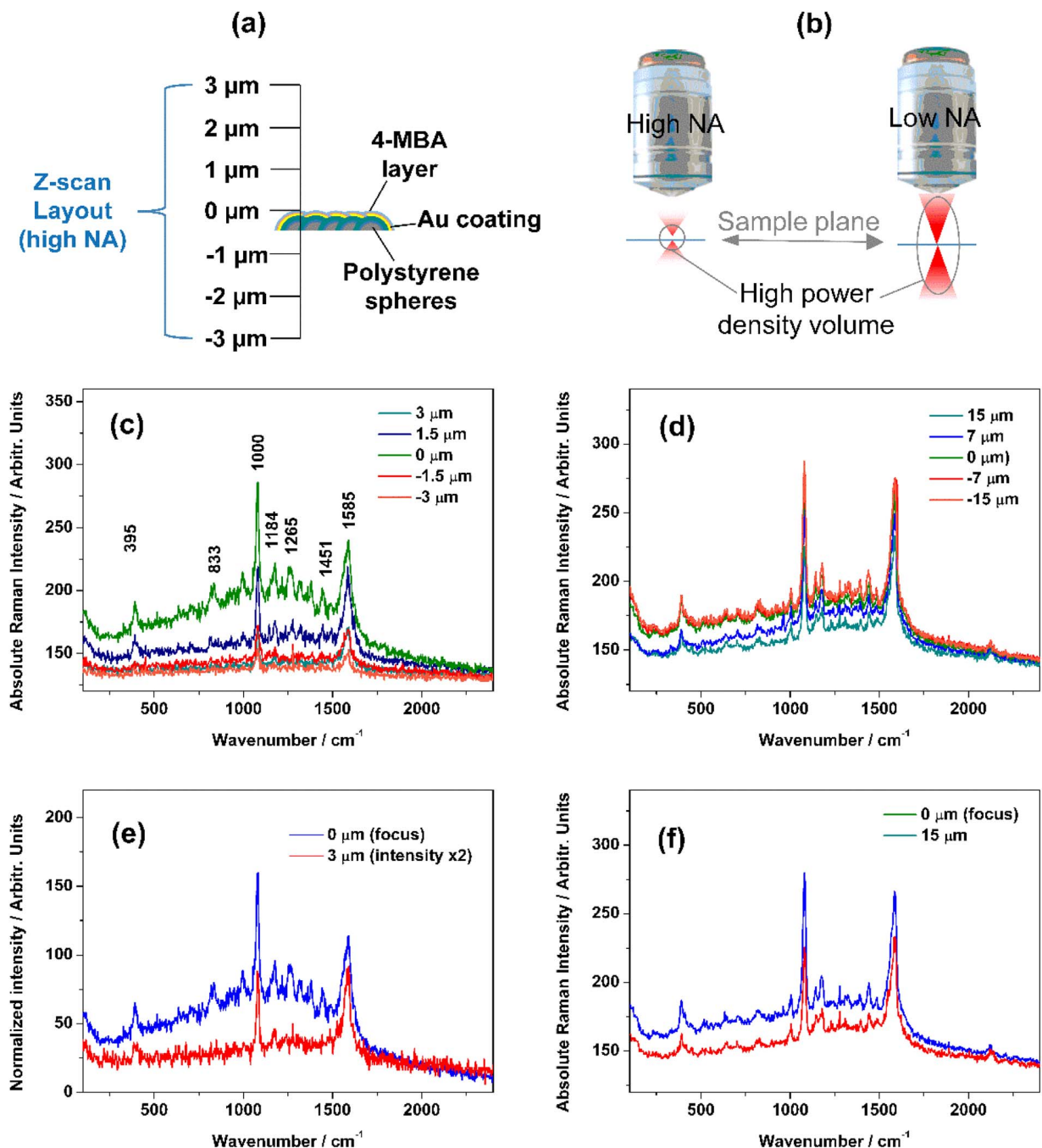

Fig. 2 (a) Schematic display of the Z-axial scan over the AuFoN substrate, with the scale indicating distances between the focus and sample surface; (b) geometrical scheme of laser beam path spread under a high-NA and a low-NA objective; representative SERS spectra recorded with (c) high NA and (d) low NA objective at different Z; (e) selected spectra from (c) at the focus ($Z = 0.12$ μm) and the first (farthest) point in the linescan, scaled to better allow visual inspection; (f) selected spectra from (d) at the focus ($Z = 0$) and the first (farthest) point in the linescan, scaled.

shown that only a tiny fraction of the entire surface covered with the adsorbed analyte actually promotes the highly enhanced SERS signal, and that the location of this high SERS enhancement regions is in between the adjacent Au half-shells.[24,25] When the focus moves vertically, these specific, well-identified locations of highest field enhancement do not change, while the magnitude of enhancement at these locations can change.

Thus, it can be assumed that when changing focus, the same collection of molecules, from the same locations respond, and not molecules located on different regions of the metal surface.

For tracking of 4-ABT signal, the background subtracted spectra were used. Hence, the area of the F1 (1080 cm$^{-1}$) at the profile peak was 5.4 times larger at the profile maximum than at the endpoints in the example presented in Fig. 3a, and on

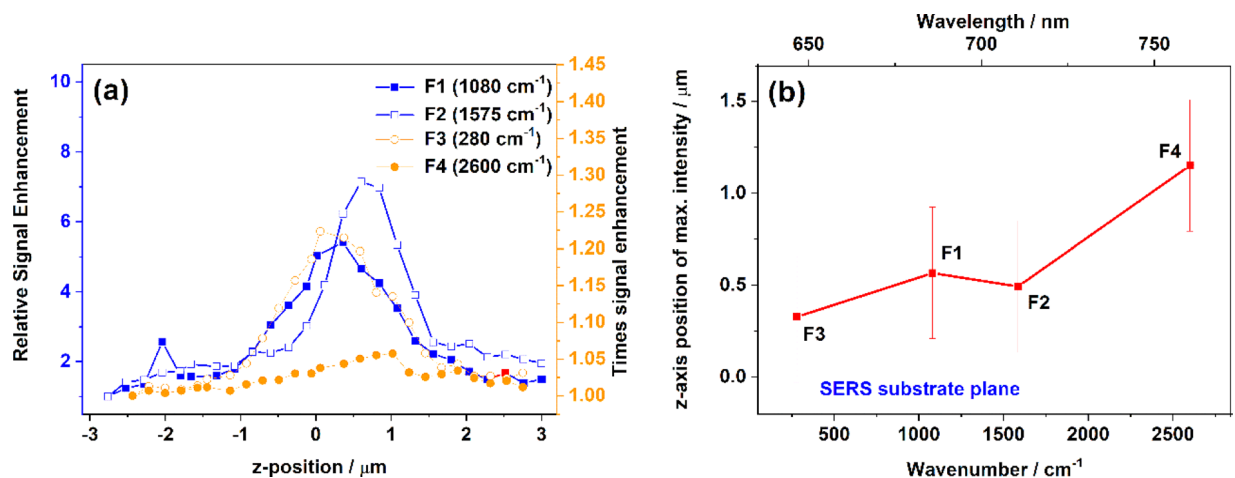

Fig. 3 (a) Signal strength profiles for different spectral regions in a typical z-linescan; (b) plot of z-positions where the profiles from (a) experience intensity maxima against their wavenumbers and wavelengths (average for 6 linescans ± S.D.).

average 5.8 ± 1.8 times (mean ± S.D.) across all acquired Z-linescans. For the F2 (1580 cm$^{-1}$) the enhancement was 7.0 and 5.5 ± 0.9 times, respectively. For the analysis of F3 and F4 which represent the background fluctuation, unprocessed spectra were used with an intact background. The F3 integrated area (around 280 cm$^{-1}$) also showed profiles similar to those of the 4-ABT bands, albeit with a lower extent of variation. Hence, F3 increased on average only 1.3 times at the profile maximum relative to the endpoints. The F4 integrated area (around 2600 cm$^{-1}$) exhibited only a small change, increasing up to 1.2 times at the profile maximum. The right vertical axis in Fig. 3a shows the times enhancement referring to the background profiles. Some outliers were noted; for instance, in one of the Z-linescans, the area enhancement at F1 maximum was as high as 12.5 times. Interestingly, this same Z-linescan featured also the highest enhancement of F3 (10.8 times), but near-average enhancement of the F2 (5.3 times), indicating that the effect was localized, and most probably induced by some SERS hot-spot. In all cases however, the intensity of the analyte bands always exhibited a greater enhancement factor than the SERS background.

It should be noted that recognizable SERS signal of 4-ABT was recorded even at the extremes of the Z-linescans, as shown in Fig. 2c and d. Judging by the small slopes of the profile tails, the main bands of 4-ABT may be recorded at even greater Z distances from the sample plane. Hence, solely geometrical treatment of the laser beam propagation along the z-direction is not appropriate. This was especially evident in experiments with laminar samples of several tens of microns thick, where signals of out-of-focus layers were recorded in both axial depth scans and lateral imaging.[26] Rather, given also the quasi-3D nature of the metallic surface, we may view the beam propagation through the "photon scattering approach" discussed in detail experimentally and theoretically by Everall *et al.*[27] and Freebody *et al.*[28] In brief, this approach considers the signal collection volume to be an extended area above and below the objective focal plane, although the probability of a scattered photon that originates outside of the focal plane reaching the detector is low, it is still non-zero.[27,28]

### 3.4. Differential signal strength profile shapes along Z

Careful observation reveals that, while the F1, F2, F3 and F4 profiles exhibit similar overall Lorentzian shapes along the Z-direction, they actually have slightly offset maxima in terms of the exact Z. In the example from Fig. 3, F1 signal strength profile peaks at 0.3 μm above the sample plane with Lorentzian FWHM of 2.0 ± 0.9 μm (mean ± S.D.), while the F2 intensity profile peaks 0.6 μm above the sample plane with Lorentzian FWHM of 2.2 ± 1.2, indicating that the two features do not experience a correlated signal strength ratio at each of the spectra acquisition points. The F3 which is closer to the laser wavelength peaks at $Z = 0.2$ μm, while F4 strength peaks at $Z = 1.3$ μm, both above the sample plane. This cannot be due to measurement errors as all the profiles are derived from the spectra within a given linescan, and each linescan exhibits a similar offset. Instead, a wavelength-dependent process is at work here.

Thus, the closer the examined wavelength is to the laser wavelength, the closer to the SERS substrate plane its signal strength profile peaks (Fig. 3b). Here, one might also suspect that the effect of chromatic aberration of the microscope objective may cause this shift, as it was shown to occur in high-detail Raman depth mapping.[26] However, this is not the case here, since, on the wavelength-scale, the difference between profiles F1 and F2 is not large enough. The exact Z where the profiles peak varies slightly from one linescan to another, but the overarching trend remains valid. Fig. 4a–d show profiles F1 and F2 from a selection of representative linescans performed at different positions on the SERS substrate, overlapped and normalized to the same minimum and maximum, to clearly show that the respective profiles differ. This excludes a constant shift in transmittance within the optical path.

Furthermore, this wavelength-dependent offset of signal enhancement profiles and their shape indicates that the ratio of main 4-ABT SERS bands is not the same for all Z axial points.

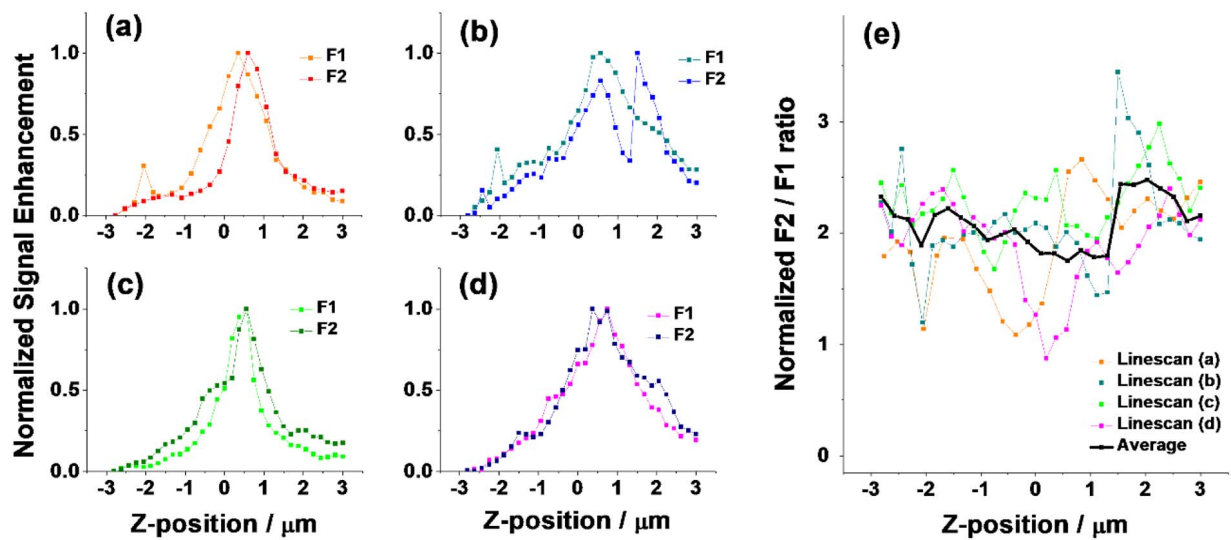

Fig. 4 (a–d) Normalized F1 and F2 profiles from multiple acquired linescans, showing the variation in profile maxima (along the $Z$-axis) and shape; (e) plot of the F2/F1 ratios calculated from the normalized profiles for individual linescans representing the integrated intensity ratio of 1580 and 1080 cm$^{-1}$ bands along acquisition points along the $Z$-axis.

Fig. 4e presents the averaged signal ratio of the 1585 and the 1080 cm$^{-1}$ bands (R = F2/F1) for the different datasets in Fig. 4a–d. Hence, it can be observed that the band ratio is the smallest at Z focus points near the sample plane, while it increases towards the linescan endpoints. This phenomenon has not yet been addressed in the literature. Because Z-linescans on SERS substrates were not previously considered in such detail, this is the first report of such a non-constant band intensity ratio. The information and results presented hitherto raise several questions with practical implications for SERS measurements: is determination of molecular orientation via band ratios a correct approach? Is the reported offset of enhancement profiles confined to AuFoN substrates or is it a wider phenomenon? Does the different laser-induced field distribution caused by defocusing change the coupling between the analyte and the substrate? After the wavelength-dependent offset of the intensity enhancement profile has been singled out in this paper, targeted studies can be designed to address the above questions. Furthermore, due to the increasing employment of statistical[29] chemometric and machine learning in the analysis of big spectral datasets,[30] such nuances have to be clarified to achieve accurate interpretation of results.

### 3.5 Structure of the SERS background

Spectra backgrounds are an integral part of the SERS spectrum, and thus should not be neglected due to its effect on overall SERS signal configuration. A study by Ebersbach et al.[31] presented a method for identifying hotspots on a SERS substrate via the ratio of background and Raman bands, demonstrating its usefulness. It seems, however, that the origin of this background is still not clarified, and probably has multiple components such as photoluminescence,[32,33] plasmon-enhanced photoluminescence,[34,35] interactions between the substrate and analyte[36] and inelastic scattering (Raman) of the substrate itself.[37,38]

There were other reports on selected properties of the background, such as its intensity modification by electric potential[39] or the effect of "hot carriers".[40] These few selected points of view present the complexity of the topic and current debates.

Although our current study does not directly address the origin of the background, we may contribute with certain observations. Correlations were previously reported between the background shape and properties of the substrate, such as the geometry of plasmonic features or LSPR wavelengths.[32,33,41] Upon visual inspection of our linescans we noted that the background changes its shape consistently depending on the position of individual spectra within the linescan. To quantitatively display this observation, we fitted the entire SERS backgrounds with Lorentzian profiles and plotted their properties in relation to the $Z$ axis. Thus, we show that the background peaks between 1100–1300 cm$^{-1}$ (∼660–800 nm) (Fig. 5a), which can be roughly linked to the reflectance minimum region of the AuFoN substrate. The intensity profile of the background peaked around the $z$-position of the SERS substrate plane (Fig. 5b). However, it appears that the background maximum changes also spectral position, reaching comparatively longer wavelengths around the substrate plane and shorter wavelengths at the defocused $z$-positions.

The shifts of the background centre occur by simply focusing, which does not imply any chemical modifications, and thus suggest that only optical/plasmonic/electromagnetic effects take place. If the locally excited plasmon resonances exhibit different spectral profiles, these could impact the spectra of the scattered photons, and thus alter the spectral profiles of the SERS spectra, including the background. Whether the main origin of the background is photoluminescence emission, as suggested by certain groups,[32–35] or inelastic scattering from the substrate,[37,38] such a plasmonic enhancement mechanism could be at work.

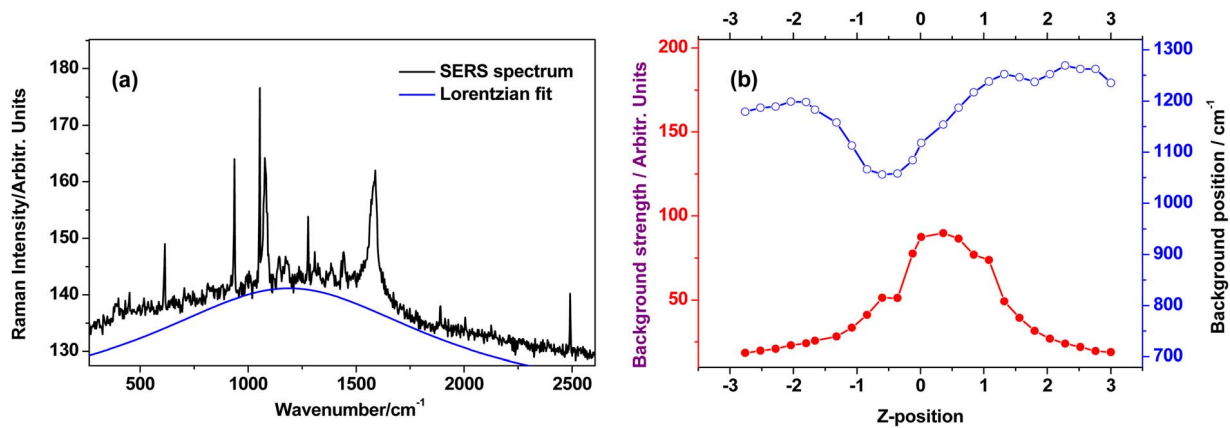

Fig. 5 Behaviour of the SERS background of the 4-ABT molecules deposited on the AuFoN substrate: (a) Lorentzian fit of the background of a representative spectrum with focus near the substrate plane in a Z-linescan; (b) intensity and wavenumber position trend of the fitted background shape across the respective Z-linescan.

### 3.6. Plasmonic effects induced by focusing/defocusing

To inquire about some possible origins of the observations discussed in previous sections, Z-dependent reflectance measurements were performed by using the high NA objective. Fig. 6 presents a series of reflectance spectra recorded at different Z positions along the vertical axis. The overall shape of the reflectance spectrum discussed already on Fig. 1b is preserved. However, slight shifts of the minimum position are observed for the different Z positions of the focus. These shifts can be understood as a manifestation in the far-field of the different distribution of the excited plasmonic modes at the near-field of the AuFoN surface.

Finally, the near-field electromagnetic response on a realistic model of the AuFoN under illumination by a focused beam was analysed by FDTD simulations. A series of simulations was performed in which the position of the focused light source relative to the plasmonic nanostructure was modified along the Z axis. This represents the equivalent of a Z-linescan in the confocal Raman experiment. Fig. 7b presents a set of electric field maps in a vertical cross-section through the AuFoN (schematized in Fig. 7a), for two different focus positions separated by 1.2 μm ($Z = +0.8$ and $-0.4$ μm), for two different wavelengths, 635 nm corresponding to the excitation laser in the experiments, and 755 nm corresponding to the F4 region at 2600 cm$^{-1}$.

Two observations can be made by analysing the electric field distribution maps: (i) the spatial distribution of the electric field near the metal surface changes when moving from 635 nm to 755 nm; this can be understood and is rather predictable by considering that the AuFoN is a plasmon resonant structure, with a wavelength-dependent optical response (see reflectivity spectrum in Fig. 1b); (ii) the electric field distribution also changes when modifying the focus position, an observation which has not been discussed previously in the context of solid SERS substrates. Calculations also show that the greatest electric fields were obtained for Z about 0.8 μm above the SERS substrate plane. This is consistent with our experimental results shown in Fig. 3a.

Corroborating these observations led to the next step in the analysis, which was to evaluate the near-field spectra at the metal surface. As such, Fig. 7c displays the normalized electric field spectra at a spatial point of high field enhancement, located near the gold surface (indicated in the inset), for the two focusing positions. The choice of this spatial point was done based on previous experimental findings on this kind of SERS substrates, which have shown that only a small fraction of the entire surface available for adsorption of the analyte actually promotes the majority of the SERS signal, and that the location of this area is in between the metallic half-shells.[24,25] The important observation in Fig. 7c is that the two near-field spectra exhibit different profiles and present maxima at different wavelengths. In turn, this has the potential to determine different wavelength-dependent responses in the SERS

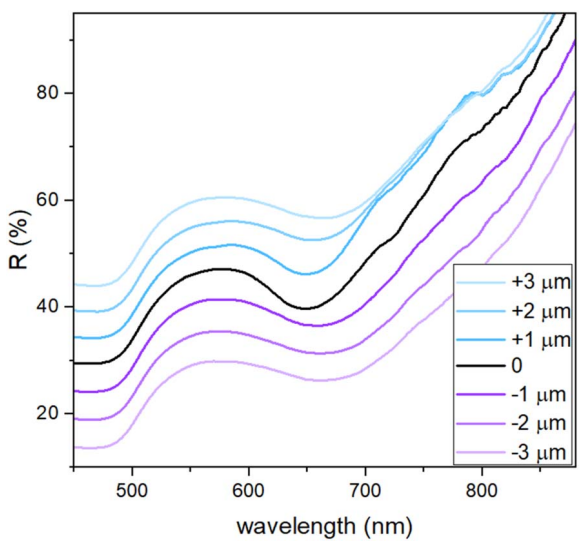

Fig. 6 Reflectance spectra recorded through the high-NA microscope objective, for different Z positions of the focus, as indicated in the legend.

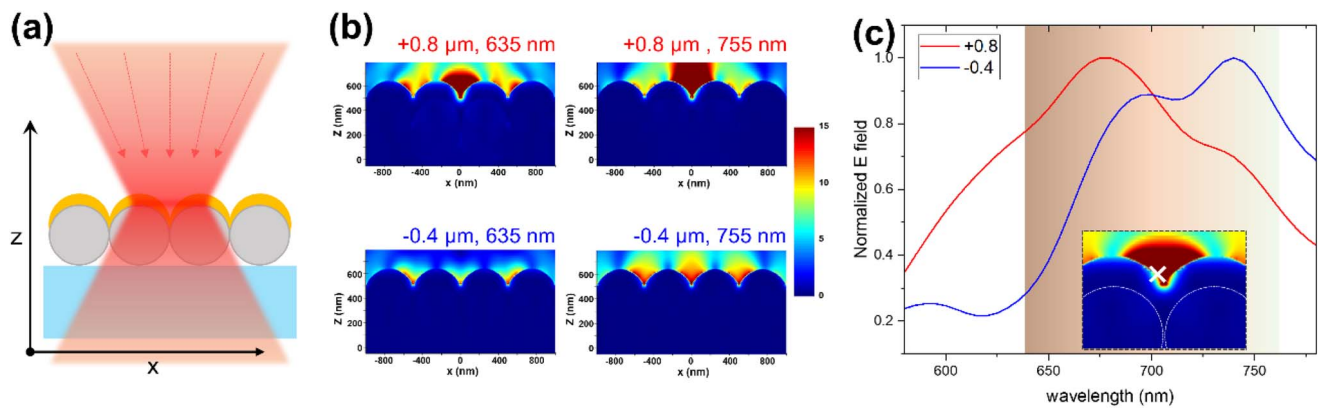

Fig. 7 (a) Schematics of the simulation setup comprising gold-coated dielectric spheres, and the focused excitation beam. (b) Electric field distribution at the surface of the AuFoN SERS substrate, at two different wavelengths (635 nm, corresponding to the excitation, and 755 nm corresponding to Raman-shifted photons at 2600 cm$^{-1}$), and for two different focus $Z$ positions (0.8 μm above and 0.4 μm below the AuFoN). (c) Electric field spectral profile near the metal surface, at the point indicated by the cross sign, for the two focus $Z$ positions.

enhancement for the two focus positions. Let's recall that the contribution of the electric field $E$ at the metal surface to the SERS enhancement, proportional to $E^4$,[28] can be understood as comprising two contributions: $E^2$ accounting for the excitation enhancement and $E^2$ for the emission (scattering) enhancement. As the SERS enhancement arises from the contributions of electric field enhancement at both the wavelength of excitation and that of Raman-shifted photons, it can be understood that different electric near-field profiles (as those observed in Fig. 7c) can induce different spectral profiles of the SERS spectra. The simulations results suggest that the effect of focusing on the SERS spectral profiles observed in our SERS analyses can be induced by the spectral distribution of the plasmon-enhanced near-fields, which can indeed be modified by focusing. The phenomenon could be linked to the angular dependence of localized plasmon modes, i.e. excitation of different components of the electric field when moving the plasmonic nanostructure out of the laser beam focus (waist).[22,23] Although to some experts in plasmonics this effect might not seem surprising, the impact on SERS induced by simply defocusing was not observed or discussed before. Therefore, more attention needs to be paid to the relation between SERS substrate morphology and SERS spectral profiles, before making complex analyses aiming, for example, to determine molecular orientation based on SERS bands intensity ratios. These observations are expected to be valid especially for SERS substrates with morphological features comparable to the excitation wavelengths, as is the case not only for metal films over nanospheres as the ones in this study, but also for many other types of SERS substrates, such as those prepared by nanoimprint lithography or photolithography.

## 4. Conclusions

Our comprehensive analysis of SERS signal intensity and spectral profiles along $Z$-axial linescans has unveiled relevant insights into the interactions between analytes and SERS substrates, specifically focusing on 4-aminobenzenethiol adsorbed onto the AuFoN plasmonic substrate. The signal strength profile demonstrates a distinctive shape along the axial $Z$-direction scans, with a frequency-dependent offset that peaks above the sample surface. Notably, the intensity ratios of various spectral regions, including SERS bands and background, display non-uniform variations along the $Z$-axis. By analysing reflectance spectra obtained through high-NA objective at different $Z$ positions, and FDTD simulations, we established that the near-field spectra reveal different profiles and maxima at varying wavelengths depending on the focus position. This correlation suggests that the trends identified in our SERS experimental data are intrinsically linked to the specific plasmonic near-field response induced by the focused beam on the SERS substrate surface. Moreover, the role of electromagnetic enhancement on scattered (emitted) photons is pivotal. Given these insights, it is essential to consider this phenomenon in future SERS analyses, as it enriches our understanding of analyte–plasmonic substrate interactions. Furthermore, comprehending the relationship between the morphology of the SERS substrate and its spectral profiles is crucial before undertaking more complex analyses, such as evaluating molecular orientation or performing quantitative assessments using SERS band intensity ratios. These findings are expected to have substantial implications for the development of SERS substrates with morphological features comparable to the excitation wavelength, paving the way for more refined and accurate applications in chemical sensing and biomedical diagnostics.

## Data availability

The data supporting this article can be found on figshare, DOI: https://doi.org/10.6084/m9.figshare.28695986.

## Conflicts of interest

There are no conflicts to declare.

## Acknowledgements

This work was supported by a grant of the Romanian Ministry of Education and Research, CNCS-UEFISCDI, project number PN-III-P4-ID-PCE-2020-1607, within PNCDI III. F. N. acknowledges financial support through the "Nucleu" Program within the National Research, Development and Innovation Plan 2022–2027, Romania, carried out with the support of MEC, project no. 27N/03.01.2023, component project code PN 23 24 01 02.

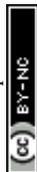